\documentclass[sigconf]{acmart}
\usepackage{multirow}
\usepackage{rotating}
\usepackage{makecell}
\usepackage{enumitem}

\AtBeginDocument{%
  }

\copyrightyear{2026}
\acmYear{2026}
\setcopyright{cc}
\setcctype{by}
\acmConference[CHI EA '26]{Extended Abstracts of the 2026 CHI Conference on Human Factors in Computing Systems}{April 13--17, 2026}{Barcelona, Spain}
\acmBooktitle{Extended Abstracts of the 2026 CHI Conference on Human Factors in Computing Systems (CHI EA '26), April 13--17, 2026, Barcelona, Spain}
\acmDOI{10.1145/3772363.3799067}
\acmISBN{979-8-4007-2281-3/2026/04}

\begin{document}

\title[Evaluating AI-Assisted Sentencing Bias Analysis for California Racial Justice Act Claims]{Can LLMs Synthesize Court-Ready Statistical Evidence? Evaluating AI-Assisted Sentencing Bias Analysis for California Racial Justice Act Claims}

\author{Aparna Komarla}
\authornote{Aparna is the founder of Redo.io (\url{https://redoio.info}), a legal AI non-profit company building technology solutions for resentencing implementation in California, US.}
\email{aparna.komarla@gmail.com}
\affiliation{
  \institution{Redo.io}
  \city{Berkeley}
  \state{California}
  \country{USA}
}


\begin{abstract}
Resentencing in California remains a complex legal challenge despite legislative reforms like the Racial Justice Act (2020), which allows defendants to challenge convictions based on statistical evidence of racial disparities in sentencing and charging. Policy implementation lags behind legislative intent, creating a 'second-chance gap' where hundreds of resentencing opportunities remain unidentified. We present Redo.io, an open-source platform that processes 95,000 prison records acquired under the California Public Records Act (CPRA) and generates court-ready statistical evidence of racial bias in sentencing for prima facie and discovery motions. We explore the design of an LLM-powered interpretive layer that synthesizes results from statistical methods like Odds Ratio, Relative Risk, and Chi-Square Tests into cohesive narratives contextualized with confidence intervals, sample sizes, and data limitations. Our evaluations comparing LLM performance to statisticians using the LLM-as-a-Judge framework suggest that AI can serve as a powerful descriptive assistant for real-time evidence generation when ethically incorporated in the analysis pipeline.
\end{abstract}



\maketitle

\section{Introduction}

A growing body of research highlights racial disparities in California's criminal legal system, particularly impacting individuals convicted of low-level and non-violent offenses \cite{peterson2016}. In 2021, nearly 20,000 individuals were incarcerated for non-violent felonies, with over 40\% serving six or more years \cite{cdcr2021}. Black individuals are incarcerated at a rate approximately 9 times greater than White individuals, driven in part by racial biases in prosecutorial discretion \cite{ppi2024, vera2024}. These disparities have been compounded by policies like the "Three Strikes and You're Out" law, which imposed life sentences for almost any crime if the defendant had two prior serious or violent convictions \cite{tsp2014}. While computational models like COMPAS (Correctional Offender Management Profiling for Alternative Sanctions) have been developed for pre-trial risk assessment, few algorithmic tools have been developed to support attorneys, self-represented litigants, and advocates in post-conviction advocacy \cite{kehl2017, propublica_machinebias}. AI-powered systems that democratize access to justice by lowering barriers to complex analyses introduce distinct human-computer interaction challenges: practitioners must be able to interrogate statistical outputs, understand their limitations, and exercise professional judgment before presenting findings in court \cite{ido_2025}.

\section{Legislative Reform History}

While California has enacted multiple retroactive sentencing reforms over the past decade, the Racial Justice Act (2020) marks a conceptual shift in resentencing policy. Earlier measures---Propositions 36 (2012), 47 (2014), and 57 (2016)---primarily addressed sentencing for property crimes, drug offenses, and parole eligibility for non-violent offenders in a deterministic fashion \cite{rand2025, cdcr2025, capol_three_strikes2025}.\footnote{Prop 36 revised California's three-strikes law to permit resentencing for qualifying persons whose third strike was not serious or violent. Prop 47 requires misdemeanor rather than felony sentencing for certain property and drug crimes and permits those previously sentenced for these reclassified crimes to petition for resentencing. Prop 57 allows certain non-violent offenders serving determinate sentences to be considered for parole after serving the full term for their primary offense.} Subsequent legislation shifted toward prosecutorial and judicial discretion: AB 2942 (2018) and AB 600 (2023) empower district attorneys and judges to directly initiate resentencing based on their own criteria \cite{capol_overview2025}.\footnote{AB 600 grants judges the authority to initiate resentencing under laws that have since been changed particularly in the absence of concurrence between district attorneys and defense counsel. AB 2942 grants district attorneys discretion to review past cases and recommend sentence reductions or recalls.} Unlike prior frameworks, the RJA introduces a novel mechanism: it permits challenges grounded in statistical evidence of systemic racial bias, requiring courts to grapple with empirical demonstrations of discrimination---a discretionary, non-deterministic standard that demands sophisticated integration of quantitative analysis with legal doctrine \cite{kuluk2024_impact, sutton2025}.

\section{Use of Statistics in Resentencing Claims}

The U.S. Supreme Court in \textit{McCleskey v. Kemp} (1987) ruled that racial bias had to be intentional and purposeful at the individual case level, effectively insulating defendants from pursuing sentencing relief when evidence of racial bias occurred at the system level \cite{kuluk2024_impact, sutton2025}. The RJA, made retroactive by AB 256, disavows this precedent and provides two statistics-based channels for defendants to address racial discrimination in charging and sentencing \cite{ebchr2023}: \textbf{(A)(3)}: The prosecution sought more severe charges against the defendant compared to other similarly situated cases in that county; \textbf{(A)(4)}: The court imposed a more severe sentence on the defendant compared to other similarly situated cases in that county. First, at the prima facie stage, a claimant must present facts indicating a "substantial likelihood" that an RJA violation occurred. Then, a discovery motion is filed for prosecutors to turn over data and information related to the claim, followed by an evidentiary hearing to establish the violation and determine eligibility for relief. For retroactive claims, incarcerated individuals may file pro se petitions using a writ of habeas corpus prior to making a prima facie showing about racial discrimination \cite{drop_lwop_guide2025}.\footnote{Habeas corpus is a legal proceeding in which an incarcerated person challenges the lawfulness of their detention or the conditions of their confinement before a court. See \url{https://selfhelp.courts.ca.gov/jcc-form/HC-001}.} Attorneys may also proactively review sentenced cases for racial injustices and pursue retroactive sentence correction for promising cases \cite{ebchr2023}. RJA filings demand expertise in both statistics and technology infrastructure to conduct large-scale analyses, the absence of which constrains the volume and pace at which claims can be made and diminishes the likelihood of their success. 

Since the passage of the RJA, some non-profit and academic research groups like the Paper Prisons Project at UC Berkeley have developed public-facing tools to assist attorneys with statistical analyses in court.\footnote{The Paper Prisons Project uses Criminal Offender Record Information (CORI) data from the California Department of Justice (DOJ). Details about the calculations in the Paper Prisons Project RJA tool are available here: \url{https://rja.paperprisons.org/}.} The researchers provide two measures on the rate at which a given charging, sentencing, or arrest-related event occurs for a selected race relative to: (a) the group's total population in the county, and (b) the total population of non-Hispanic White individuals in the county. The tool is limited for several reasons: (1) disproportionate representation in carceral settings does not automatically imply a causal relationship between race and sentencing bias; (2) users cannot apply controls on criminal activity outside of the sentencing outcome of focus, such as prior prison sentences that help standardize a group for apples-to-apples comparisons \cite{bazelon2025, actionable_disparity2023}. Furthermore, the tool does not provide interpretive guidance for practitioners without statistical training, leaving attorneys to bridge the gap between disparity metrics and legal argumentation on their own.

\subsection{RJA Implementation Gaps}

The use of statistical evidence in arguments related to discrimination in criminal law represents a significant departure from existing practice and introduces distinct interpretive challenges for courts and practitioners \cite{sutton2025, bazelon2025, actionable_disparity2023}. Analyses of RJA implementation across California suggest that attorneys have limited guidance on conducting empirical studies, identifying similarly situated defendants, and proving causal relationships between race and sentencing outcomes \cite{sutton2025, bazelon2025, actionable_disparity2023}. Because these claims are labor-intensive and the law is still developing, statistics-based RJA claims have moved slowly, taking months or even years of litigation and multiple trips to higher courts \cite{crpc2025, actionable_disparity2023}. The Committee on Revision of the Penal Code (CRPC) found only three cases where a trial court reached the merits of a statistical claim after an evidentiary hearing, suggesting that the RJA remains significantly underutilized \cite{crpc2025}.\footnote{Two of these three cases were denied because the courts rejected the expert witness' methodology and faulted the defense for failing to provide examples of people charged differently for similar conduct.} Tools that democratize legal research and accelerate case evaluation are especially powerful for public defender offices and self-represented litigants with limited resources to hire statisticians and build complex databases \cite{actionable_disparity2023, sutton2025, chien2020}. Disparities in funding between district attorneys and public defenders, the absence of institutional public defenders in nearly half of the counties in the state, and the emergence of private contracting in public defense create a troubling irony: the defendants most harmed by racial bias in sentencing---and thus most likely to benefit from RJA relief---are the least able to access the expertise and infrastructure needed to prove their claims \cite{wright2024, greene2025, lao2022}. 

\subsubsection{Research Question}
Our work aims to answer three primary questions in socio-technical system design: \textbf{RQ1.} How do we design AI systems that effectively translate between legal requirements and statistical reasoning in the RJA? \textbf{RQ2.} How accurately do LLMs synthesize evidence about racial bias in sentencing while considering all contextual factors? \textbf{RQ3.} How should AI-generated evidence reports be presented to support legal argumentation that is both ethical and effective given that flawed LLM outputs can perpetuate deeper systemic racial harms?

\section{AI Assistant for Bias Analysis and Court-Ready Evidence}

We address critical limitations in existing tools for RJA implementation by treating race as a treatment effect to identify disparities between White and non-White sentenced individuals, while enabling attorneys to control for prior prison commitments and define similarity thresholds appropriate to their case \cite{gaebler2022causal, clair2016, kuluk2024_impact}. Recent benchmarks like QRDATA on LLM performance for statistical and causal reasoning tasks show that general-purpose models achieve accuracy of 58\% and frequently miscalculate p-values \cite{llm_stats_2021, razeghi_2022_pval} (see Appendix F). Statistical problem-solving requires precise numerical computation, deep conceptual understanding, and logical derivation---all of which differ markedly from the pattern-matching strengths of pretrained LLMs \cite{bommasani_2022}. Given these limitations of LLMs, we deliberately conduct statistical analysis using traditional programming methods, reserving AI for downstream tasks. LLMs are employed solely for evidence synthesis and narrative construction, serving as an interpretive bridge between technical analysis and legal argumentation. The challenges with using AI to generate such narratives are multifold. For example, it is imperative for an attorney to contextualize observed disparities or lack thereof with confidence intervals, sample size, and underlying data limitations in public records disclosure (see Appendix D and E). The system is therefore designed to augment, not replace, attorney judgment. Users retain control over cohort definitions, variable selection, and the decision to submit any generated output to a court. The LLM serves as a drafting assistant whose outputs require professional review before use in legal proceedings.

\subsection{System in Action}
To illustrate, consider an attorney seeking to demonstrate that Black individuals in Contra Costa County were oversentenced for firearm enhancements during robberies. Using our platform, the attorney constructs a similarly situated cohort through (1) query-based filters, where current commitments contain PC12022 (firearm enhancement) and PC211 or PC212 (robberies)\footnote{"PC" refers to the California Penal Code.}; sentencing county is Contra Costa; or (2) vector-based searches.\footnote{Platform tools: similarly situated case search (\url{https://tool.redoio.info/scenario_builder} and \url{https://tool.redoio.info/similar_records}), bias analysis toolkit (\url{https://tool.redoio.info/bias_analysis}). Cases are represented as vectors in an n-dimensional space, where each dimension captures a sentencing-relevant metric such as offense severity or degree of violence. Similarity measures including Cosine, Tanimoto, and Euclidean distance are then used to identify comparable cases.} The attorney specifies a reference ethnic group (non-Hispanic White) and applies statistical methods including Odds Ratios, Relative Risk, and Chi-Square Test---established as valid benchmarks in prior RJA cases such as \textit{People v. Windom}---to measure disparities in sentencing \cite{ball2025, bazelon2025, kuluk2024_impact}. Once computations are complete, the platform generates an evidence report for this Contra Costa County scenario by prompting GPT-5-mini with the computed results (point estimates, 95\% confidence intervals, p-values, contingency table cells, and sample sizes), ethnicity labels, the sentencing outcome under comparison, and a reference context of research papers on RJA proceedings.\footnote{Full prompt inputs and reference context are described in Appendix B and C. Source materials are available at \url{https://github.com/redoio/ai_evals}.} The prompt instructs the model to synthesize findings across methods, flag limitations in the underlying data, attribute observed disparities to systemic factors rather than inherent criminal behavior, and return a cohesive report (see Appendix H).

Since the RJA does not explicitly prescribe which confounding variables to control for, the platform allows users to define selection criteria (e.g., excluding cases with prior super strike or violent felony commitments) when constructing cohorts. Attorneys can then rapidly iterate on analyses under varying definitions of similarly situated cohorts and thresholds for biased sentences, incorporating additional controls in response to judicial or opposing counsel feedback.

\begin{table}[h!]
\centering
\small
\caption{Contingency Table Structure and Formulas for Racial 
Bias Analysis.}
\label{tab:or_rr_formulas}
\begin{minipage}{\columnwidth}
\centering

\begin{tabular}{@{}l cc c@{}}
\toprule
 & \textbf{Event} & \textbf{No Event} & \textbf{Total} \\
\midrule
Non-White & A & B & $A + B$ \\
White & C & D & $C + D$ \\
\midrule
Total & $A + C$ & $B + D$ & $N$ \\
\bottomrule
\end{tabular}

\medskip

\begin{tabular}{@{}l l@{}}
\toprule
\textbf{Measure} & \textbf{Formula} \\
\midrule
Relative Risk (RR) & $\dfrac{A/(A+B)}{C/(C+D)}$ \\[10pt]
Odds Ratio (OR) & $\dfrac{A \times D}{B \times C}$ \\[10pt]
Chi-Square ($\chi^2$) & $\dfrac{N(AD-BC)^2}{(A+B)(C+D)(A+C)(B+D)}$ \\[6pt]
\bottomrule
\end{tabular}

\medskip

\begin{tabular}{@{}l @{\hspace{4pt}} l @{\hspace{4pt}} l@{}}
\toprule
\textbf{Measure} & \textbf{Value} & \textbf{Meaning} \\
\midrule
\multirow{3}{*}{RR / OR} & $= 1$ & No racial disparity \\
 & $> 1$ & Non-White overrepresented \\
 & $< 1$ & White overrepresented \\
\addlinespace[4pt]
\multirow{2}{*}{$\chi^2$} & $p < 0.05$ & Significant association \\
 & $p \geq 0.05$ & No significant association \\
\bottomrule
\end{tabular}
\end{minipage}
\end{table}

\begin{figure}
\includegraphics[width=\columnwidth, alt={Process of synthesizing statistical findings and generating an evidence report.}]{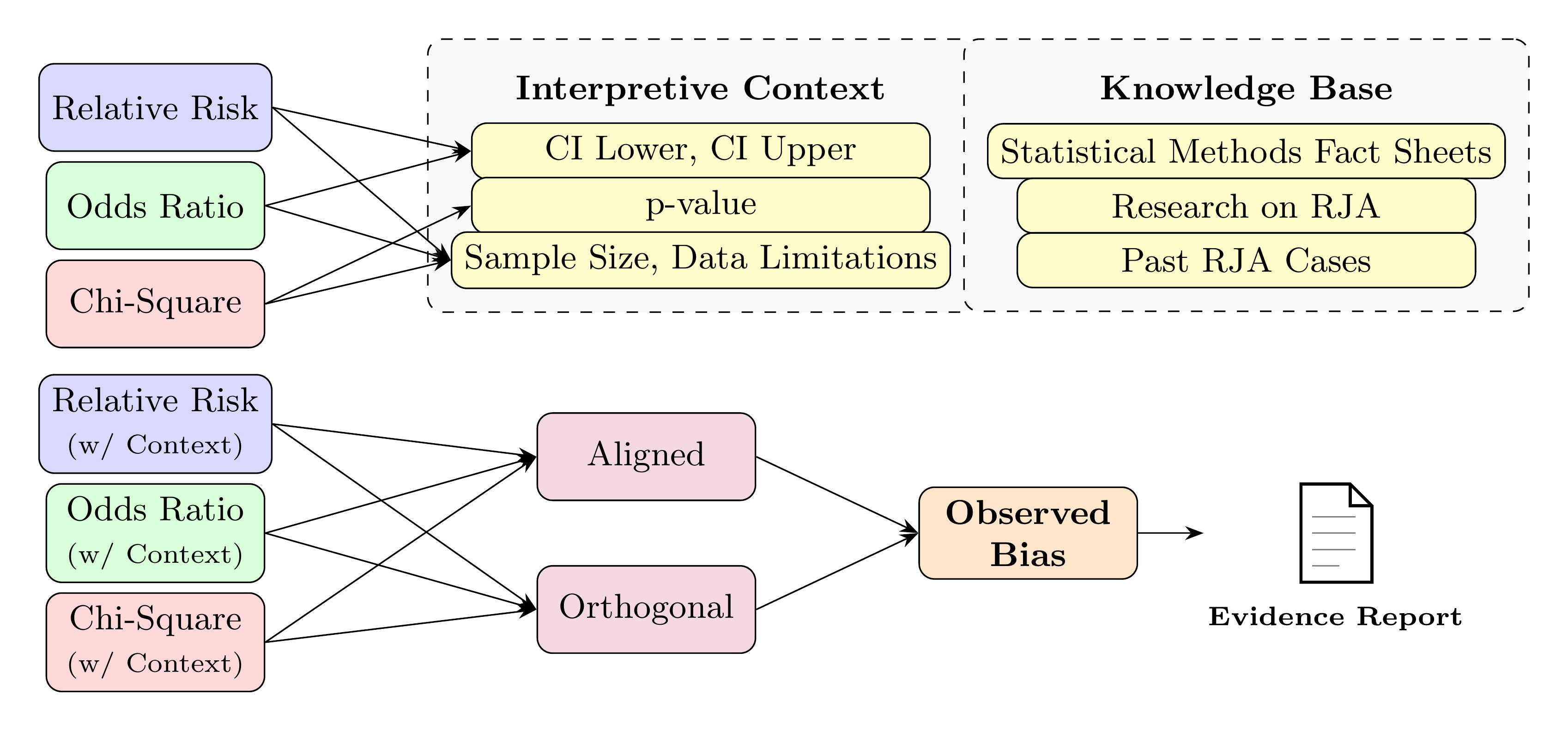}
\caption{Process of synthesizing statistical findings and generating an evidence report. See Appendix H for a sample report.}
\Description[Process of synthesizing statistical findings.]{Process of synthesizing statistical findings and generating an evidence report. See Appendix H for a sample report.}
\label{fig:bias_pipeline}
\end{figure}

\subsubsection{Value Added from AI Analysis}
Without LLMs, one might employ rule-based templates to synthesize findings into a narrative. However, each statistical method introduces multiple binary contextual dimensions (e.g., CI width, inclusion of 1, sample size adequacy, data limitations), and the three methods can align or diverge in $2^3 = 8$ distinct patterns. The resulting combinatorial space requires $2^{14} = 16{,}384$ unique templates---a number that grows exponentially with each additional method (see Appendix G).\footnote{For instance, adding logistic regression with four binary dimensions would increase the total to $2^{14} \times 2^4 \times 2 = 2^{19} = 524{,}288$ templates.} Beyond scale, rule-based approaches impose rigid cutoff thresholds that fail to capture nuances between statistical methods. LLMs obviate explicit template enumeration by synthesizing evidence in a context-sensitive manner, distinguishing between close and clear convergence or divergence across methods---distinctions that would require extensive conditional logic under a rules-based framework. By reducing the overhead of preparing a prima facie showing or discovery motion, AI tools can increase both the volume and pace of RJA filings.

\begin{figure}[h!]
\centering
\includegraphics[width=0.4\textwidth, alt={Description of figure 8}]{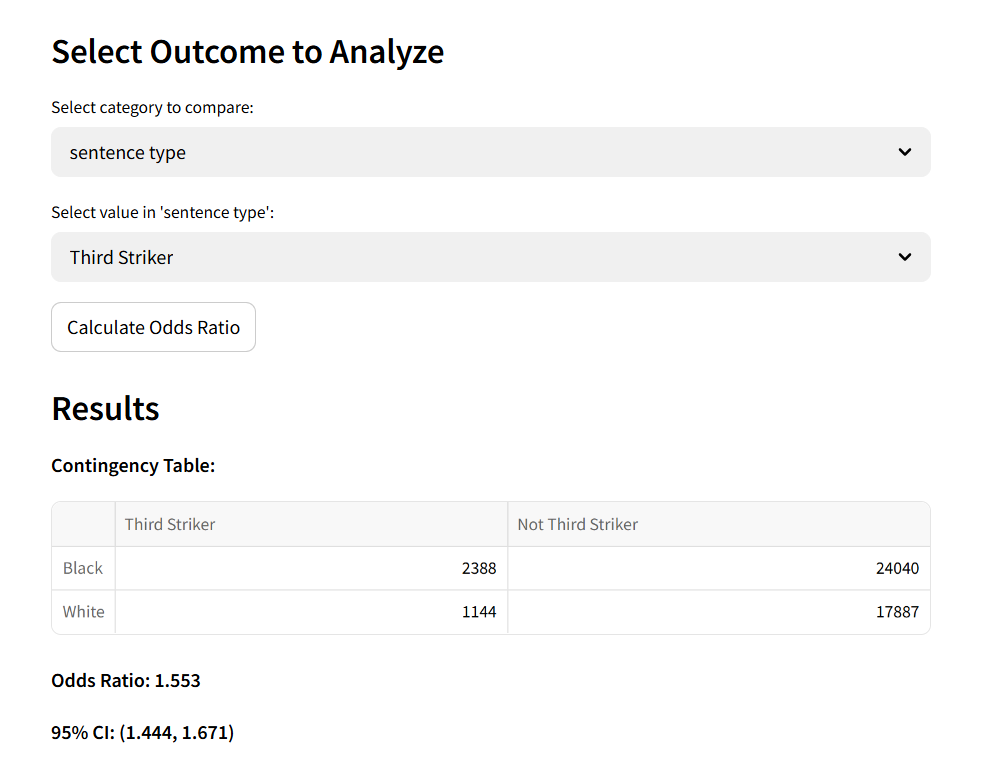}
\includegraphics[width=0.4\textwidth, alt={Description of figure 9}]{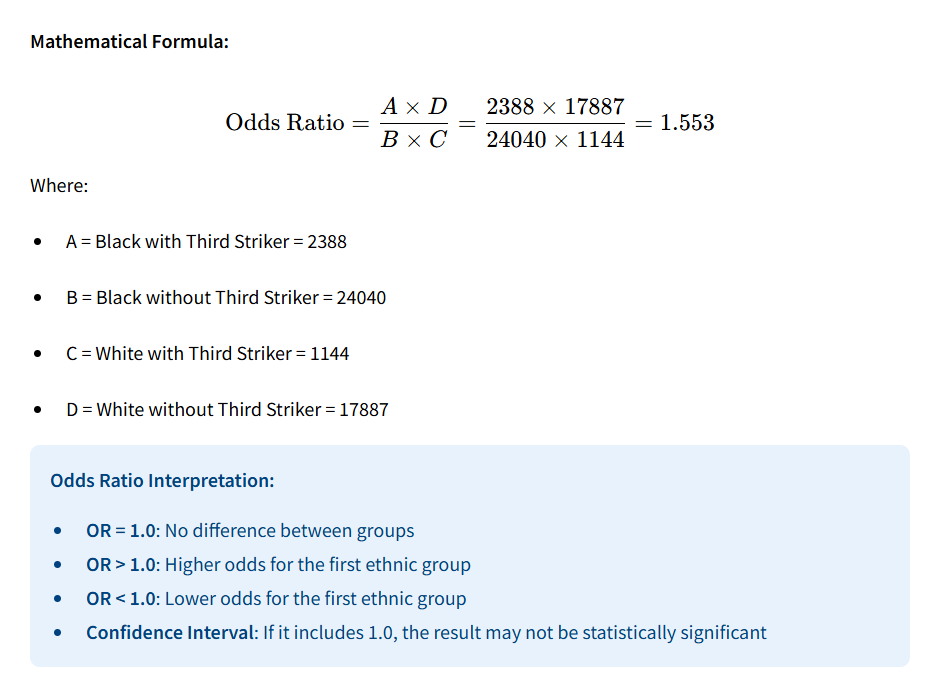}
\caption{View of \url{https://tool.redoio.info/bias_analysis} to evaluate the strength of the association between demographics and sentencing outcomes (e.g. 'Third Striker'). See Appendix G for an example of the bias analysis report.}
\Description[View of \url{https://tool.redoio.info/bias_analysis}]{View of \url{https://tool.redoio.info/bias_analysis} to evaluate the strength of the association between demographics and sentencing outcomes (e.g. 'Third Striker'). See Appendix G for an example of the bias analysis report.}
\label{fig:decision_assistant_a}
\end{figure}

\begin{table*}[h]
\centering
\caption{RJA Statistical Analysis: Traditional Process vs. Redo.io AI Platform}
\label{tab:rja_comparison}
\small
\renewcommand{\arraystretch}{1.3}
\begin{tabular}{p{4cm}p{5cm}p{5cm}}
\hline
\textbf{Task} & \textbf{Traditional Process} & \textbf{Redo.io AI Platform} \\
\hline
Data acquisition & CPRA request or Data Sharing Agreement with CDCR (weeks to months) & Immediate access (pre-obtained from CDCR) \\
Data scientist & Contract with vendor (\$\$\$) & Not required (free platform) \\
Cohort generation & Manual analysis (hours to days) & Automated (30 minutes) \\
Bias analysis report & Custom development (\$\$\$, days) & Automated (30 minutes) \\
\hline
\textbf{Total cost} & \textbf{Thousands of dollars} & \textbf{Free} \\
\textbf{Total time} & \textbf{Weeks to months} & \textbf{1 hour} \\
\hline
\end{tabular}
\end{table*}

\section{Implementation and Results}

\subsection{Evaluation of AI Assistant for Bias Analysis}
We evaluate GPT-5-mini's ability to accurately interpret statistical results and synthesize observed biases into court-ready evidence reports.\footnote{GPT-5-mini (\url{https://developers.openai.com/api/docs/models/gpt-5-mini}), medium reasoning, accessed via direct API calls. Inputs consist of concatenated instruction-context prompts with a maximum output length of 20,000 tokens.} We generated 30 reports by varying sentencing county, sentencing outcome, ethnicity, and controls on prior record with three statistical methods: Odds Ratio, Relative Risk, and Chi-Square Test (see Appendix G for a sample report). We employ two evaluation approaches: human evaluators with a minimum of upper-division undergraduate training in statistics, and an LLM-as-a-Judge. Both receive the same inputs (evidence reports, raw statistical findings, and evaluation rubric) and score each dimension on a 0--1 scale.\footnote{For LLM-judgment, to avoid reusing cached responses across evaluation runs, each evaluation request is sent with a unique value for both user and prompt\_cache\_key (e.g. a new UUID per request). That way each request is routed to a different cache bucket and the API does not return a cached result from a previous run. The interpretation text and criteria may be identical; only the routing parameters differ. We run 15 evaluations per report and dimension to compute the mean and standard deviation of each score.}  Recent work demonstrates that LLM-based evaluation correlates strongly with human judgments and captures nuanced aspects of reasoning quality that lexical similarity metrics miss, making this approach well-suited for domain-specific tasks such as statistical interpretation \cite{gu2024_llmjudge, bavaresco_llm_judge_2025, yancey_2023_rating, nagarkar2025}.

\subsubsection{Bias Analysis Rubric}
We test six primary dimensions of AI-powered evidence reports related to reasoning and accuracy, and assess known limitations of LLM judges such as their sensitivity to perturbations in inputs (see Appendix C for the full evaluation rubric) \cite{llm_stats_2021, llm_stats_2024, llm_stats_2025, llm_stats_2026, nagarkar2025, shi2025, bordt2024}:

\begin{enumerate}
\item \textbf{Confidence Interval Contextualization:} Ability to contextualize statistical methods with their confidence intervals (CI), including both the width and the inclusion or exclusion of 1. A result with a wide CI that includes 1 is generally considered imprecise and lacking statistical significance, which an evidence report must flag to avoid overstating findings in court.
\item \textbf{P-Value Contextualization:} Ability to contextualize statistical methods such as the Chi-Square Test with their p-values. A result with a p-value greater than 0.05 is typically considered lacking statistical significance at $\alpha = 0.05$, and the report should clearly communicate this threshold to attorneys.
\item \textbf{Dataset Limitations:} Ability to contextualize statistical results with limitations of the underlying dataset, such as populations or offense categories excluded due to privacy restrictions in public records law. For instance, CDCR does not report cases of juvenile offenders under the CPRA to prevent re-identification and harms to personal privacy or safety.\footnote{The California Public Records Act (CPRA) grants the public the right to access records held by state and local government agencies, subject to specific exemptions for privacy and law enforcement. See \url{https://firstamendmentcoalition.org/handbook/california-public-records-act/}. The California Department of Corrections and Rehabilitation (CDCR) is the state's system of over 30 prisons.} Our dataset also excludes charges that did not result in conviction, misdemeanor convictions, and other cases that did not result in a prison sentence. Observed disparities should therefore be interpreted with the caveat that the record of criminal activity captured in the dataset is incomplete.
\item \textbf{Sample Size and Balance Ratio:} Ability to contextualize results with sample size, balance ratio, and representativeness of both ethnic groups. A well-balanced analysis generally requires at least a 60-40 split between two groups to ensure sufficient cases to speak to sentencing patterns in each.
\item \textbf{Appropriate Attribution:} Observed disparities in sentencing must be attributed to systemic factors, not inherent criminal behavior of any racial group. Evidence reports should not imply that disparities reflect a ground truth about propensity to commit crime. This criterion is especially critical given the sensitivity of RJA proceedings and the known potential for algorithmic systems in criminal justice to encode and amplify racial biases \cite{propublica_machinebias, rudin2020_secrecy, rudin2019_blackbox, miron2020}. AI-generated language that fails to maintain this distinction risks reinforcing the very narratives that the RJA was enacted to challenge.
\item \textbf{Method Comparison:} Ability to compare results from different statistical methods, especially when they yield divergent conclusions, and present a cohesive interpretation. For example, Odds Ratio and Relative Risk answer related but distinct questions: Odds Ratio compares the odds of an outcome between groups, while Relative Risk compares probabilities directly. The two measures converge when the outcome is rare (prevalence below approximately $10\%$), but diverge as the outcome becomes more common, with Odds Ratio tending to overestimate the magnitude of association compared to Relative Risk. This divergence is not an error but a known statistical property that an evidence report must explain to avoid misleading a court.
\end{enumerate}

\begin{table}[t]
\caption{Difference (D) between Avg. Machine (L) and Human (H) Evaluation Scores (30 Cases)}
\label{tab:average_scores}
\centering
\small
\begin{tabular}{@{}lcccc@{}}
\toprule
\textbf{Evaluation Dimension} & \textbf{L} & \textbf{L*} & \textbf{H} & \textbf{D$^\dagger$} \\
\midrule
OR CI Score (OR CI$_{\text{score}}$)                         & 0.79 & 0.06 & 0.80 & $-$0.01 \\
RR CI Score (RR CI$_{\text{score}}$)                         & 0.76 & 0.09 & 0.80 & $-$0.04 \\
OR Sample Size Score (OR Smp$_{\text{score}}$)               & 0.49 & 0.08 & 0.61 & $-$0.12 \\
RR Sample Size Score (RR Smp$_{\text{score}}$)               & 0.49 & 0.08 & 0.61 & $-$0.12 \\
Cross-Method Comparison Score (Cmp$_{\text{score}}$)         & 0.69 & 0.11 & 0.78 & $-$0.09 \\
Chi-square Sample Size Score ($\chi^2$ Smp$_{\text{score}}$) & 0.50 & 0.09 & 0.59 & $-$0.09 \\
p-Value Contextualization Score ($p_{\text{score}}$)         & 0.89 & 0.05 & 0.84 & +0.05  \\
Limitations Score (Lim$_{\text{score}}$)                     & 0.75 & 0.01 & 0.94 & $-$0.19 \\
Attribution Score (Att$_{\text{score}}$)                     & 1.00 & 0.00 & 1.00 &    0.00 \\
\midrule
\textbf{Overall Score (Ovr$_{\text{score}}$)} & \textbf{0.71} & \textbf{0.03} & \textbf{0.76} & \textbf{$-$0.05} \\
\bottomrule
\end{tabular}

\smallskip
\begin{minipage}{\columnwidth}
\raggedright\footnotesize
*Mean standard deviation ($\bar{\sigma}$) across 15 LLM evaluation runs per report (see Appendix K for details). $^\dagger$Plus (+) indicates the LLM judge scored higher than human evaluators.

\smallskip
\scriptsize
OR/RR CI\textsubscript{score}: Contextualization with CI width and inclusion/exclusion of 1.0 \\
OR/RR Smp\textsubscript{score}: Contextualization with sample size and group representativeness \\
Cmp\textsubscript{score}: Reconciliation across methods especially when they diverge \\
$\chi^2$ Smp\textsubscript{score}: Chi-square with sample size considerations \\
$p$\textsubscript{score}: Significance ($p < 0.05$ or $p \geq 0.05$) and contingency table \\
Lim\textsubscript{score}: Dataset limitations and impact on outcomes \\
Att\textsubscript{score}: Attribution to systemic factors, not inherent criminal behavior \\
Ovr\textsubscript{score}: Average of all scores
\end{minipage}
\end{table}

\begin{table*}
\caption{LLM-as-a-Judge Evaluation Results (Selected Reports and Metrics; \textbf{Eth. 1} = White)}
\label{tab:evaluation}
\centering
\small
\begin{tabular}{@{}llccccccccccc@{}}
\toprule
\textbf{Eth. 2} & \textbf{Outcome} & \textbf{OR (95\% CI)$_{\text{raw}}$} & \textbf{$\chi^2_{\text{raw}}$} & \textbf{$p_{\text{raw}}$} & \textbf{OR CI$_{\text{score}}$} & \textbf{Lim$_{\text{score}}$} & \textbf{OR Smp$_{\text{score}}$} & \textbf{Att$_{\text{score}}$} & \textbf{Cmp$_{\text{score}}$} & \textbf{$p_{\text{score}}$} & \textbf{Ovr$_{\text{score}}$} \\
\midrule
Black & Drug Crimes & 0.48 (0.09, 2.38) & 0.32 & 0.5733 & 1 & 1 & 0.75 & 0.75 & 1 & 1 & 0.81 \\
Mexican & Property Crimes & 1.89 (0.78, 4.55) & 1.52 & 0.2172 & 1 & 1 & 0.75 & 1 & 0.75 & 1 & 0.89 \\
\bottomrule
\end{tabular}
\end{table*}

\subsection{Results}
Overall, AI-powered evidence generation performs well (LLM: 0.71, human: 0.76) on a 0--1 scale, though with meaningful variation across dimensions that carries practical implications for legal practitioners. LLM evidence synthesis is strongest on qualitative safeguards. The model appears to reliably attribute observed sentencing disparities to systemic factors rather than inherent criminal behavior (\textbf{Attribution}: 1.00 for LLM and human). The model also performs well on acknowledging data constraints, though this dimension produces the largest evaluator discrepancy (\textbf{Dataset Limitations}: LLM 0.75, human 0.94), suggesting that human evaluators are more generous in crediting the model's treatment of these issues. The most consistent weakness lies in \textbf{Sample Size Contextualization}, where scores across all statistical methods remain below 0.50 for the LLM judge and below 0.61 for human evaluators. The model frequently reports group counts but fails to contextualize findings with respect to balance ratios, group representativeness, or what small samples imply for the reliability of point estimates. Evidence synthesis also struggles with \textbf{Cross-Method Comparison} (LLM: 0.69, human: 0.78), and the high variability in LLM evaluations of the same report ($\bar{\sigma}$ = 0.11) indicates that the model's ability to reconcile divergent statistical findings is both limited and inconsistent.\footnote{For each report, we compute $\sigma$ across 15 independent LLM evaluations of the \textbf{Cross-Method Comparison} dimension. $\bar{\sigma}$ is the average of these standard deviations across all 30 reports. See Appendix J and K for details.} Odds Ratio and Relative Risk, for instance, converge when outcomes are rare but diverge substantially otherwise, and explaining this divergence requires statistical reasoning that current LLMs appear to handle unreliably. These results suggest that AI-generated evidence reports can accelerate the initial stages of RJA case preparation, particularly prima facie showings, but necessitate attorney review before court submission---especially regarding sample size adequacy and cross-method reconciliation.

\section{Limitations}

\subsection{Observational Study Constraints}
As with many observational studies on discrimination, our analysis faces well-documented methodological challenges \cite{gaebler2022causal}. Defining causal estimands when the treatment variable is an immutable characteristic such as race remains conceptually unresolved. Estimates may also suffer from omitted-variable bias if users do not adjust for all relevant covariates, including those absent from our dataset, or from post-treatment bias when adjusting for covariates determined downstream of the treatment variable (in our case, ethnicity). These issues are compounded by the multi-stage nature of the criminal legal system. Sentencing disparities reflect accumulated biases across stops, arrests, and charging decisions, which are difficult to disentangle from one another \cite{guarnera2024bias}. This sequential structure also introduces selection bias: treatment and control groups may exclude individuals who could have received the sentencing outcome after charging but were never charged, or who were diverted from charging itself at earlier stages of the justice system \cite{gaebler2022causal}. Long criminal histories further complicate analysis, as prior convictions simultaneously reflect past systemic biases and influence current sentencing decisions, making it difficult to isolate the effect of race at any single stage \cite{kurlychek2019cumulative, kutateladze2014cumulative}.

\subsection{Evaluation Constraints}
Both human and LLM judges exhibit known biases and are sensitive to perturbations in framing and rubric design \cite{chen-etal-2024-humans}. These concerns are compounded when assessment rubrics may themselves be influenced by the model outputs \cite{shankar2025}. Moreover, prior work has shown that LLM judges tend to favor their own outputs, a phenomenon known as self-preference bias \cite{llm_judge_self_bias_1, llm_judge_self_bias_2, llm_judge_self_bias_3}. Because we use GPT-5-mini for both report generation and evaluation, this bias is a potential concern. However, our results do not show a clear pattern of self-leniency: the LLM judge scores lower than human evaluators across most dimensions (overall gap of $-$0.05), with \textbf{p-Value Contextualization} as the sole exception (LLM: 0.89, human: 0.84). Conversely, the large gap on \textbf{Dataset Limitations} (LLM: 0.75, human: 0.94) suggests the LLM judge may be stricter than human evaluators on certain qualitative dimensions. In future work, we plan to expand our evaluations to other general-purpose LLMs, with and without fine-tuning on statistical reasoning tasks, to reduce dependence on a single model for both report generation and evaluation.

\section{Discussion} 

\subsection{Statistical Significance and Legal Standards}
This work highlights a fundamental tension between statistical inference and legal argumentation. The RJA's "significant difference" standard invites an interpretation that may not align with how statisticians understand significance. In 2016, the American Statistical Association issued a statement cautioning that scientific conclusions and high-stakes decisions should not be based on whether a p-value passes a specific threshold alone \cite{actionable_disparity2023}. Court judgments of whether a disparity's magnitude qualifies as meaningful difference under the RJA are still evolving, and clear precedents remain scarce. Courts will also have to decide how to weigh magnitude against uncertainty: is a small but precisely estimated disparity more concerning than a larger but noisier one? As these legal standards develop, AI tools must evolve accordingly, adapting not only to new statistical benchmarks but also to emerging judicial interpretations of what constitutes actionable evidence \cite{actionable_disparity2023}.

\subsection{Design Implications for AI in High-Stakes Professional Workflows}
Our experience highlights several design tensions relevant to AI tools in professional settings beyond law. The interpretive layer must be transparent about the boundaries of its reasoning: practitioners need to understand not just what the model concluded, but why it weighed evidence in a particular way (e.g., current LLMs struggle to present conflicting signals without resolving them prematurely). The same model properties that make LLMs effective drafting assistants (fluency, confidence, coherence) can also make their errors harder to detect, reinforcing the need for evaluation frameworks that go beyond surface-level quality.

\section{Future Work}
While prior research underscores the importance of AI augmenting rather than replacing legal practice, effectively bridging statistical analysis and courtroom presentation remains an open challenge \cite{ido_2025, li_2024, yu_2024, yuan_2021, zhang_2024}. We identify three priorities for future work. First, we plan to investigate deployment safeguards: what human oversight is necessary before AI-generated evidence reports are used in court, and how can practitioners validate outputs and identify errors prior to submission? Second, we aim to conduct a user study examining how attorneys interact with evidence reports, including whether the current report presentation format supports effective legal reasoning or introduces new risks of misinterpretation. Third, we intend to expand our evaluation framework to assess model performance on edge cases, such as small sample sizes, highly imbalanced cohorts, and scenarios where multiple statistical methods yield contradictory conclusions.

\section{Conclusion}
This work demonstrates that AI-powered tools can meaningfully accelerate the initial stages of case preparation under the RJA. Our platform enables attorneys to construct similarly situated cohorts, detect racial disparities through multiple statistical methods, and generate court-ready evidence reports---capabilities that accelerate preparation for prima facie and discovery motions. Our evaluation of GPT-5-mini's evidence synthesis shows that the model performs reliably on qualitative safeguards such as attribution and dataset limitation awareness, while revealing room for improvement in reconciling divergent outputs from different statistical methods. These findings contribute to a growing body of work on LLM reasoning in high-stakes domains, suggesting that LLMs are better suited as interpretive bridges between statistical analysis and legal argumentation than as autonomous statistical reasoners. More broadly, this work highlights a fundamental tension that will shape the future of AI in legal proceedings: the complex interplay between statistical inference and legal standards of evidence. As courts develop precedents for what constitutes a "significant difference" under the RJA, the tools that support this litigation must evolve in parallel---not only improving in technical accuracy, but also adapting to the judicial and evidentiary standards that ultimately determine whether statistical findings translate into injustices that warrant relief.

\begin{acks}
We thank Susan Champion (Deputy Director, Three Strikes Project at Stanford Law) and Caneel Fraser (Director, Indigent Defense Improvement Division at the Office of the State Public Defender) for their guidance. We acknowledge Beyond AI Cloud for their contribution to developing Redo.io's data platform. We also thank Walter Campbell (Principal Research Associate at the Urban Institute) for his contributions to our research and tool design. Lastly, we acknowledge Responsible AI in Legal Services (RAILS) at Duke University's School of Law for their legal research support. This work is partly funded by Microsoft and the Urban Institute through the Catalyst Grant program.
\end{acks}
\bibliographystyle{ACM-Reference-Format}
\bibliography{sample-base}

\appendix

\section{GenAI Use Disclosure}

Authors leveraged Claude and GPT to assist with generating LaTeX code for tables, and grammar and cohesion checks.

\section{Useful Links}
General information about Redo.io: \url{https://redoio.info}\\
Catalyst Grant program funding: \url{https://www.urban.org/projects/catalyst-grant-program/redoio}\\
GitHub: \url{https://www.github.com/redoio}\\
AI platform: \url{https://tool.redoio.info}\\
Bias analysis toolkit: \url{https://tool.redoio.info/bias_analysis}\\
Vector similarity search: \url{https://tool.redoio.info/similar_records}\\
Simple data analysis: \url{https://tool.redoio.info/data_analysis}

\subsection{Data Sources}
Open datasets: \url{https://www.github.com/redoio/offenses_data} and \url{https://data.world/redoio}\\
Data dictionary: \url{https://tool.redoio.info/data_dictionary}
Frequently Asked Questions (FAQs): \url{https://tool.redoio.info/faqs}

\section{Evaluation Rubrics \& Results}

Rubric for evaluating AI-powered evidence analysis:
\url{https://www.github.com/redoio/ai_evals/blob/main/evaluation_rubric.md}\\
LLM-as-a-Judge evaluation results: \url{https://www.github.com/redoio/ai_evals/blob/main/evaluation_results.xlsx}\\
Human evaluators' results: \url{https://www.github.com/redoio/ai_evals/blob/main/evaluation_results_human.xlsx}

\section{Dataset Architecture}

Our publicly accessible database comprises approximately 95,000 individuals incarcerated in CDCR, obtained through CPRA requests sent to CDCR every 6-8 months (current versions are as of December 2023 and March 2025). The database consists of three relational tables with 75 variables joined by de-identified CDCR IDs:

\begin{itemize}
\item \textbf{Demographics:} Ethnicity, controlling offense, offense dates, parole eligibility, institution, sentence length, sentencing county, and sentence type (third-striker, second-striker, life with/without parole, determinate).
\item \textbf{Current Commitments:} Offense code, enhancements, sentence type, court-determined length, and sentencing county for each active commitment.
\item \textbf{Prior Commitments:} Same structure as current commitments plus release dates for completed terms.
\end{itemize}

All CDCR IDs are de-identified using MD5 hashing with a protected seed value to prevent external identification while maintaining cross-table linkage.

\section{Dataset Constraints}

By leveraging the CPRA, we access prison sentencing data without establishing formal Data Sharing Agreements (DSAs) with government agencies.\footnote{A Data Sharing Agreement is a legally binding contract between a public agency such as CDCR and another institution, such as a public defender's office, district attorney's office, or academic law clinic, that governs the terms of data transfer, storage, permissible use, and retention.} Since the CPRA permits the dissemination of data and analyses derived from public records, this approach facilitates smoother integration with public-facing tools and mitigates privacy concerns when interfacing with LLM APIs. Data accessed under a DSA is typically more detailed and may include personally identifiable information, but DSAs impose stricter constraints on how the data can be used, shared, and stored, limiting the development of public-facing applications. Establishing such agreements may also be infeasible in counties without institutional public defense offices. Conversely, the CPRA's privacy restrictions exclude juvenile identification and health information, making it challenging to examine cases for mitigating factors and exceptional conduct in prison.

Our prison sentencing data obtained under the CPRA represent a subset of the records that CDCR typically provides under a formal Data Sharing Agreement (DSA) between a county agency and the state. DSAs for prison sentencing data commonly include variables such as sex registration status, mental health level of care, California Static Risk Assessment (CSRA) scores, COMPAS scores, milestones achieved, education credits, and gender. Although our CPRA requests to CDCR enumerate all of these variables and others, CDCR has stated that it cannot disclose them under the CPRA due to privacy restrictions in existing law. For example, CDCR has indicated that providing gender and sex registration status would violate the Prison Rape Elimination Act (PREA). Separately, CDCR has stated that it is unable to provide records pertaining to specific milestone and rehabilitation credits, citing Penal Code sections 11075, 11076, and 13102, as well as Government Code section 7927.705, on the grounds that these data are classified as Criminal Offender Record Information (CORI). This classification is notable because milestone and rehabilitation credit data do not compromise individual privacy, HIPAA protections, or personal safety, yet their designation as CORI effectively shields them from public disclosure and limits the ability of researchers and advocates to assess rehabilitation outcomes alongside sentencing patterns.

Charging data, which is essential for examining racial disparities under RJA A(3) motions, presents a separate challenge: it is not held by CDCR but rather distributed across individual county agencies, including police departments and district attorneys' offices, making systematic collection difficult due to the decentralized nature of these records.

\section{Sample Query in QRDATA Benchmark}

Liu et al. constructed a benchmark QRDATA with 411 questions along with 195 data sheets from open-source textbooks, online learning resources, and academic papers that are accompanied by data. The statistical reasoning questions in QRDATA are precisely the kind of outcomes that we seek to offer attorneys. Example of a \textbf{Scenario-Question-Answer} triad in the dataset:
\begin{quote}
\textbf{Scenario:} A migraine is a particularly painful type of headache, which patients sometimes wish to treat with acupuncture. To determine whether acupuncture relieves migraine pain, researchers conducted a randomized controlled study where 89 females diagnosed with migraine headaches were randomly assigned to one of two groups: treatment or control. 43 patients in the treatment group received acupuncture that is specifically designed to treat migraines. 46 patients in the control group received placebo acupuncture. 24 hours after patients received acupuncture, they were asked if they were pain free. The research data is in the file migraine.csv.\\
\textbf{Question:} "In which group did a higher percentage of patients become pain free 24 hours after receiving acupuncture?" Please answer with "treatment group" or "control group."\\
\textbf{Answer:} "Treatment group"
\end{quote}

\section{Rules-Based Template Calculation}

\textbf{Step 1: Individual Method Templates.}
Each method has binary dimensions that produce independent outcomes:

\begin{table}[h!]
\centering
\small
\begin{tabular}{p{1.2cm}p{4.5cm}cc}
\toprule
\textbf{Method} & \textbf{Dimensions} & \textbf{Calc.} & \textbf{Templates} \\
\midrule
Odds Ratio & CI width, CI inclusion of 1, Sample size, Data limitations & $2^4$ & 16 \\
Relative Risk & CI width, CI inclusion of 1, Sample size, Data limitations & $2^4$ & 16 \\
Chi-Square & p-value, Sample size, Data limitations & $2^3$ & 8 \\
\bottomrule
\end{tabular}
\caption{Individual method templates based on binary dimensions.}
\label{tab:method_templates}
\end{table}

\noindent
\textbf{Step 2: Combined Method Templates}
All possible combinations across three methods:
\begin{equation}
2^4 \times 2^4 \times 2^3 = 2^{11} = 2048
\end{equation}

\noindent
\textbf{Step 3: Alignment Patterns.}
Each method produces a directional result: Bias (B) or No Bias (N). With 3 methods, each having 2 possible outcomes:
\begin{equation}
\text{Patterns} = 2^3 = 8
\end{equation}

\begin{table}[h!]
\centering
\small
\begin{tabular}{cccp{2.5cm}}
\toprule
\textbf{OR} & \textbf{RR} & \textbf{Chi} & \textbf{Description} \\
\midrule
B & B & B & All aligned (bias) \\
N & N & N & All aligned (no bias) \\
B & B & N & OR \& RR aligned \\
N & N & B & OR \& RR aligned \\
B & N & B & OR \& Chi aligned \\
N & B & N & OR \& Chi aligned \\
N & B & B & RR \& Chi aligned \\
B & N & N & RR \& Chi aligned \\
\bottomrule
\end{tabular}
\caption{Alignment patterns for three methods.}
\label{tab:alignment}
\end{table}

\noindent
\textbf{Step 4: Total Templates.}
\begin{equation}
\text{Total} = 2^{11} \times 2^3 = 2^{14} = 16,384
\end{equation}

\section{Example of AI-Powered Evidence Report}

Example of a report generated using \url{https://tool.redoio.info/bias_analysis}:
\begin{quote}
\textbf{Executive Summary}
The sentencing outcome analyzed is sentence type "DSL" and the comparison is between White and Hispanic individuals. There is evidence of a sentencing disparity that disproportionately results in Hispanic individuals receiving DSL; the disparity is moderate in magnitude and warrants further scrutiny of systemic sentencing processes.

\textbf{Findings}\\
\textit{What statistical methods were used?}\\
The odds ratio was used to compare the odds that White individuals receive DSL relative to Hispanic individuals; it expresses the ratio of odds of the outcome across the two groups. The relative risk was used to compare the probability (risk) of receiving DSL for White individuals relative to Hispanic individuals; it provides an intuitive measure of how much more or less likely the outcome is in one group versus the other. The chi-square test of independence was used to evaluate whether an association exists between ethnic group and sentencing outcome, testing whether the observed distribution of cases across groups departs from what would be expected under independence.

\textit{What meaning and trends did we find?}\\
The odds ratio of 0.521 with 95\% confidence interval (0.364, 0.745) indicates that the odds of White individuals receiving DSL are lower than the odds for Hispanic individuals, and the confidence interval excludes 1.0, indicating statistical significance at the conventional 0.05 level. The relative risk of 0.69 with 95\% confidence interval (0.566, 0.857) indicates that the risk of receiving DSL is lower for White individuals compared to Hispanic individuals, and the confidence interval excludes 1.0, indicating statistical significance. The chi-square test produced a statistic of 12.26 with p-value 0.0005, providing strong evidence of an association between ethnic group and the DSL sentencing outcome that is unlikely to be due to chance. Taken together, the three methods are concordant: they show a statistically significant disparity in DSL sentencing that disfavors Hispanic individuals; the odds ratio shows a larger numerical contrast than the relative risk (a pattern expected when the outcome is relatively common), and all measures point to a consistent direction and statistically significant magnitude of disparity in sentencing outcomes.

\textit{What ethical considerations are necessary when interpreting this result?}\\
Observed disparities in sentencing outcomes should be interpreted as evidence of differences in how sentences are imposed rather than as evidence of inherent differences in criminal propensity among ethnic groups. Explanations for disparities should emphasize systemic factors, such as prosecutorial discretion, implicit bias, differential charging practices, and structural inequalities, rather than characteristics of the groups themselves. Care must be taken to avoid attributing causation to race without further case-level analysis and to ensure that interpretations focus on sentencing disparities alone.

\textbf{Analytical Constraints and Considerations}\\
\textit{What is the input dataset unable to tell us and why?}
The input dataset does not by itself explain case-level drivers of sentencing decisions such as offense severity, plea bargaining details, prosecution charging rationales, judge-level discretion, or validated measures of prior criminal history; absence of these covariates limits ability to isolate whether observed disparities are attributable to race or to other case-specific factors. The data are anonymized and may lack certain variables excluded under disclosure rules, which prevents follow-up verification of individual case contexts and constrains the ability to control for potentially confounding factors. Statistical association observed in this aggregated data cannot establish causal attribution and may be influenced by measurement error, missing information, or systemic processes earlier in the criminal legal pipeline.

\textbf{Key Terms and Methodology}\\
The odds ratio is a ratio comparing the odds of an event occurring in one group to the odds in another group; it is commonly used to quantify relative differences in odds between two groups. Relative risk is the ratio of the probability of an event in one group to the probability in another group and conveys how much more or less likely the event is in one group compared with the other. The chi-square test is a statistical test that assesses whether two categorical variables are associated by comparing observed cell counts to expected counts under independence. A confidence interval is a range of values, derived from the data, that is believed to contain the true population parameter with a specified level of confidence. A p-value is the probability of observing data as extreme as those observed, or more so, if the null hypothesis of no association were true. The balance ratio is a descriptive measure indicating the relative sample sizes or proportions between comparison groups and is used to assess whether groups are similar in size for comparative analysis.

Read the full report here: \url{https://github.com/redoio/ai_evals/tree/main/report_001}

\end{quote}

\section{Example Evaluation Outputs}

\subsection{OR CI Contextualization}
\subsubsection{LLM Judge (Score: 0.75/1)}
The explanation correctly interprets that the odds ratio is not statistically significant because its 95\% confidence interval spans values below and above 1.0, and it discusses uncertainty/precision in general (noting limited power and wide confidence intervals can reduce ability to detect effects) and explains the implications of a CI that includes 1.0. However, it fails the essential requirement of explicitly stating the odds ratio CI numeric values (0.663, 2.308).
\subsubsection{Human Judge (Score: 0.75/1)}
The report clearly mention the odds ratio and its confidence interval, It clearly states the confidence interval includes 1 and its corresponding explanation of it being non significant. It also mention wide intervals and imprecision. It further connects uncertainty to weak evidence.

\subsection{Sample Size Contextualization}
\subsubsection{LLM Judge (Score: 0.5/1)}
The explanation only minimally addresses sample-size issues. It generically notes that "event counts are small" and that "statistical power and precision are constrained when event counts are small," which acknowledges how small counts affect precision. However, it does not explicitly state the sample size or contingency-table counts, does not assess whether the available sample is adequate for reliable analysis, and fails to mention the required minimum count of cases for evaluation.
\subsubsection{Human Judge (Score: 0.5/1)}
Although the report acknowledged the small event number and imprecision, it does not explicitly mention the sample size and does not mention whether it's adequate for reliable analysis (15 cases). It does not clearly distinguish between overall sample size and event counts.

\subsection{Cross-Method Comparison}
\subsubsection{LLM Judge (Score: 0.5/1)}
The explanation covers several important elements: it defines odds ratio and relative risk conceptually, states that the OR and RR point estimates point in the same direction, and explicitly notes that both confidence intervals include 1.0 and that the chi-square p-value is not significant, tying the chi-square result to the lack of significance in the OR/RR CIs. However, it fails to explicitly report or compare the numeric OR and RR.
\subsubsection{Human Judge (Score: 0.5/1)}
The report compares the odds ratio, relative risk, and chi square test, and mention that they align in direction and are statistically significant. But it fails to mention the conceptual differences and the magnitude quantification beyond general descriptions.

\section{LLM-as-a-Judge Evaluation Summary Across 30 Reports (15 Runs per Report)}

\begin{figure}[h]
\centering
\includegraphics[width=\columnwidth]{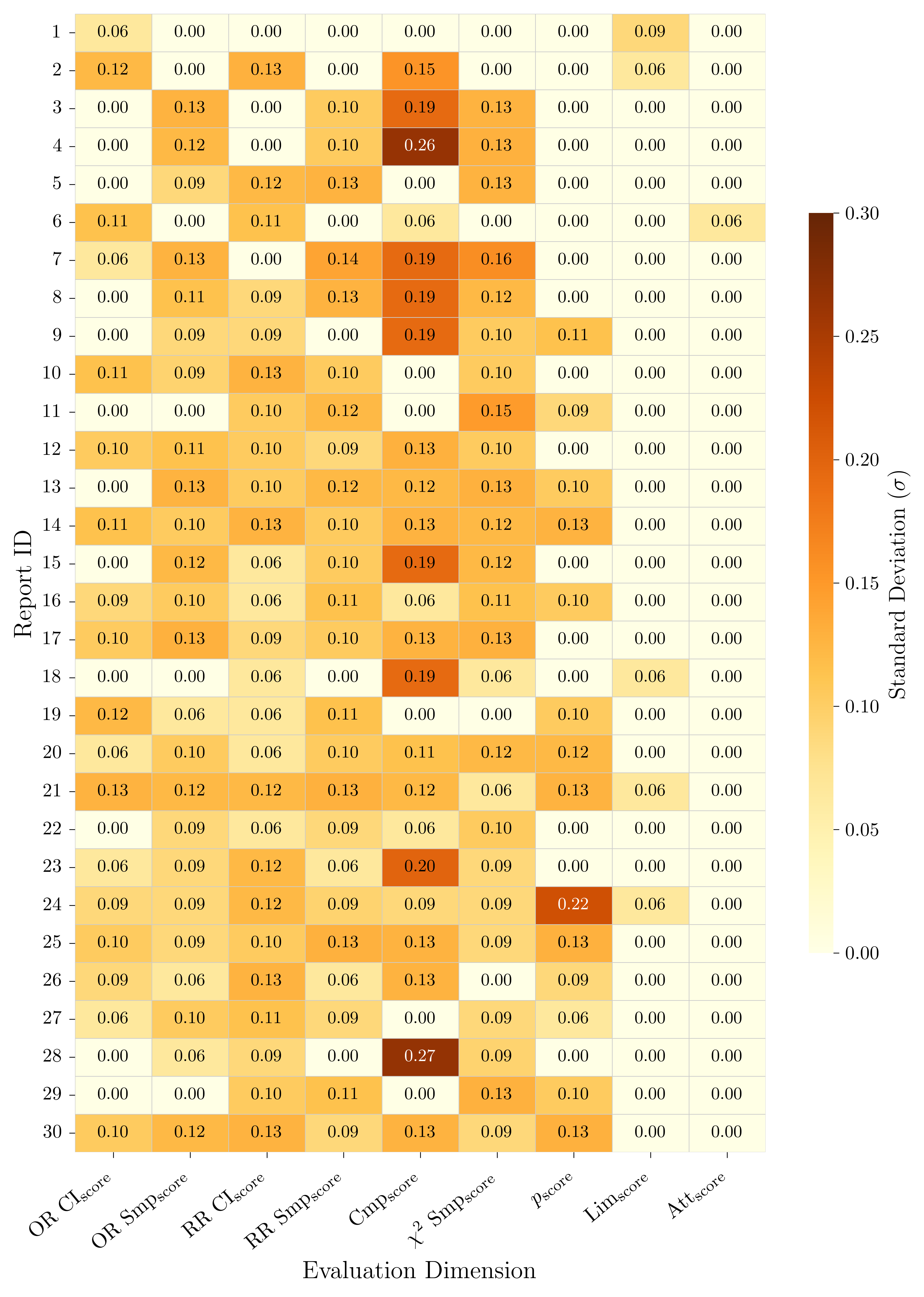}
\caption{Standard deviation of LLM-as-a-Judge scores across 15 evaluation runs per report. Higher values (darker cells) indicate greater scoring variability. Cross-Method Comparison (Cmp) exhibits the highest variability, while Limitations (Lim) and Attribution (Att) are nearly constant.}
\Description[Standard deviation of LLM-as-a-Judge scores]{Standard deviation of LLM-as-a-Judge scores across 15 evaluation runs per report. Higher values (darker cells) indicate greater scoring variability. Cross-Method Comparison (Cmp) exhibits the highest variability, while Limitations (Lim) and Attribution (Att) are nearly constant.}
\label{fig:std_heatmap}
\end{figure}

\begin{table}[h]
\caption{LLM-as-a-Judge Evaluation Summary Across 30 Reports (15 Runs per Report)}
\label{tab:eval_summary}
\centering
\small
\begin{tabular}{@{}lcccc@{}}
\toprule
\textbf{Evaluation Dimension} & \textbf{Mean} & \textbf{Std} & \textbf{Min} & \textbf{Max} \\
\midrule
OR CI Score (OR CI$_{\text{score}}$) & 0.79 & 0.06 & 0.33 & 1.00 \\
RR CI Score (RR CI$_{\text{score}}$) & 0.76 & 0.09 & 0.37 & 1.00 \\
OR Sample Size Score (OR Smp$_{\text{score}}$) & 0.49 & 0.08 & 0.00 & 0.75 \\
RR Sample Size Score (RR Smp$_{\text{score}}$) & 0.49 & 0.08 & 0.00 & 0.75 \\
Cross-Method Comparison Score (Cmp$_{\text{score}}$) & 0.69 & 0.11 & 0.58 & 0.75 \\
Chi-square Sample Size Score ($\chi^2$ Smp$_{\text{score}}$) & 0.50 & 0.09 & 0.00 & 0.75 \\
p-Value Contextualization Score ($p_{\text{score}}$) & 0.89 & 0.05 & 0.63 & 1.00 \\
Limitations Score (Lim$_{\text{score}}$) & 0.75 & 0.01 & 0.72 & 0.75 \\
Attribution Score (Att$_{\text{score}}$) & 1.00 & 0.00 & 0.98 & 1.00 \\
\midrule
\textbf{Overall Score (Ovr$_{\text{score}}$)} & \textbf{0.71} & \textbf{0.03} & \textbf{0.42} & \textbf{0.84} \\
\bottomrule
\end{tabular}

\smallskip
\footnotesize
\noindent\textbf{Note:} Mean and standard deviation are computed across 30 reports, each evaluated 15 times. All scores range from 0.00 to 1.00.
\end{table}

\end{document}